\documentclass[9pt]{article}
\usepackage{latexsym}
\usepackage{graphicx}
\usepackage{amsfonts,amssymb}

\begin{document}

\title{The Role of Weight on Community Structure of Networks}
\author{Ying Fan$^{1}$, Menghui Li$^{1}$,
Peng Zhang$^{1}$, Jinshan Wu$^{2}$\footnote{Author for correspondence: jinshanw@phas.ubc.ca}, Zengru Di$^{1}$\\
\\\emph{ 1. Department of Systems Science, School of Management,}\\
\emph{Beijing Normal University, Beijing 100875, P.R.China}
\\\emph{2. Department of Physics \& Astronomy, University of British Columbia, }\\ \emph{Vancouver, B.C. Canada, V6T 1Z1}}

\maketitle

\begin{abstract}
The role of weight on the weighted networks is investigated by
studying the effect of weight on community structures. We use
weighted modularity $Q^w$ to evaluate the partitions and Weighted
Extremal Optimization algorithm to detect communities. Starting
from idealized and empirical weighted networks, the distribution
or matching between weights and edges are disturbed. Using
dissimilarity function $D$ to distinguish the difference between
community structures, it is found that the redistribution of
weights does strongly affect the community structure especially in
dense networks. This indicates that the community structure in
networks is a suitable property to reflect the role of weight.
\end{abstract}



\section{Introduction}{\label{correlat}}

Link Weights, as strength of the interaction represented by
networks, are believed to be an important variable in networks. It
gives more information about networks besides its topology
properties dominated by links. Recently more and more study in
complex networks focus on the weighted networks. The problems
involve the definition of weight and other quantities which
characterize the weighted networks\cite{6,8,Fan}, the empirical
studies of its statistical properties\cite{10,11,12,Li}, evolving
models\cite{7,9,16,17,18,zhou}, and transportation or other
dynamics on weighted networks\cite{13,14,15,zhou1}.

However, how important is the weight, or what significant changes
on network structures are induced when weight is changed? This, we
call, the role of weight, should be a fundamental question in the
study of weighted networks. But it has not been investigated
deeply in the previous studies.

The role of weight should be first investigated by analyzing the
correlation between edge weight and other properties. In this way,
it attempt to answer the question that if there is some internal
mechanism strongly determining weights. For example, one may image
the edge betweenness effects edge weight largely because the
larger edge betweenness implies the edge has the more important
role in communication on networks so that the weight on the edge
might be also larger. If this is true, the weight should be less
important for the networks because the weight is dependent on
links or the weight is determined by the network topology.

For the database of our scientific collaboration networks of
econophysicists, BNU-email network, and monkey's societies, based on
the standard method for linear correlation analysis in mathematical
statistics, we get the correlation results for the edge weights and
edge betweenness, the edge weight and the sum of degrees of its two
ends vertices. The results are shown in Table 1. All the
coefficients of correlation for links are less than 0.25. These
negative results reveal that the weight is really an independent
variable for complex networks. By the way, we also get the
correlation coefficients for the vertex weight (strength of vertex)
and its degree. They are 0.79, 0.44. and 0.71 for our above three
networks. The results are rational because they are all contributed
by the connected links.

\begin{center}
Table 1. Correlation coefficients for weight and other quantities
\end{center}
\begin{center}
\begin{tabular}{|c|c|c|c|c|}
\hline

\multicolumn{2}{|c|}{} & EP-SCN & BNU-Email & Monkey \\
\hline
 & Weight-Betweenness & 0.0055 & 0.028 & 0.19  \\ \cline{2-5}
\raisebox{1.7ex}[0pt]{Links}& Weight-Degree of ends & 0.226 & 0.028 &  0.14  \\
\hline
Vertices & Strength-Degree & 0.79 & 0.44 & 0.71 \\
\hline
\end{tabular} \label{cluster}
\end{center}

From the above negative conclusion on correlation analysis, we
know the weight is an independent variable at some level. This
even makes the work on the role of weight more attractive: since
it's somehow independent, how significant is it?

The effects of weight on the network structures can be
investigated on two classes of properties: single vertex
statistics and correlation statistics. The former includes vertex
based properties such as degree, clustering coefficient, and the
later includes global properties such as distance, betweenness and
especially community structure. Except we consider the difference
of above properties between unweighted and weighted networks, an
important way to study the effects of weight is to consider their
difference after we disturb the weight distribution. A natural way
to disturb weights is to redistribute weights onto different
edges\cite{Li}, where we either change the distribution of weight
or mix the matching between weights and edges. However, our
previous investigation shows that this redistribution has little
effect on the single vertex statistics, neither significant effect
on distance. This negative conclusion looks like weight does not
have significant effect on network structures. However, we
strongly suspect that it's only because we have not found the
proper measurement to present its effect.

An analogy between networks and condensed matter may give us some
clues for insightful investigation. In condensed matter, at most
cases, an effective single electron picture is well enough for a
large number of phenomena. An effective field is used to represent
effect from all electrons and lattice ions in condensed matter.
However, there are something beyond this single particle scheme so
that it requires to consider the correlation between electrons.
Single vertex statistics naturally belong to the former class,
even possible including some global properties such as distance.
But community structure measures directly the correlation among
vertices. Therefore, in this work, we use the community structure
as a measurement on the role of weight.

In binary networks, the community structure is defined as groups
of network vertices, within groups there are dense internal links
among nodes, but between groups nodes loosely connected to the
rest of the network\cite{Girvan}. However, as we have mentioned
above, link weight is an independent variable and should have some
important effects on structure and function of networks. As for
community structures, the definition of the community must
integrate links with link weights. Newman has generalized the
modularity $Q$ to weighted modularity $Q^w$ for evaluation the
partitions of weighted networks\cite{8}:
\begin{equation}
Q^w=\frac{1}{2w}\sum_{ij}[w_{ij}-\frac{w_iw_j}{2w}]\delta(c_i,c_j),
\label{weightmodularity}
\end{equation}
It takes both links and link weights into account. Usually, groups
separated with the link weights should be different from the
result based only on topological linkage. Given the same
topological structure, different assignments of link weights may
result in different community structures. Our basic question is
how will the community structure change when the weights are
disturbed.

There are several questions should be answered before the
realization of the above ideas. First, what are the networks for
this investigation. Our previous analysis uses networks of
Econophysicists as our typical networks\cite{Fan}. Recently, we have
got more datas on BNU-email networks and monkey's
societies\cite{8,Zhang}. Hopefully, dense weighted networks will
give us more confirmative conclusions. Besides these real networks,
we can also construct idealized ad hoc weighted networks for our
investigation. Second, how to extract and to evaluate community
structure from a given network. Currently, there are several typical
algorithms in the literature: Hierarchical Clustering, betweenness
based GN algorithm\cite{Mixing}, Potts model based
algorithm\cite{Jorg}, Extremal Optimization algorithm\cite{EO}, and
so on. We have generalized several approaches to weighted networks
and investigated their performance\cite{RevWeight}. Here, we use
only Weighted Extremal Optimization (WEO) algorithm. The approach of
WEO is directly related with the definition of weighted modularity
$Q^w$. It performs well in weighted networks. Then the third, how to
compare different community structures among the same set of
vertices. We have proposed dissimilarity function $D$ in
\cite{Zhang} to measure the difference between partitions. Starting
from two community structures $\left\{A_{1},A_{2},\dots\right\}$ and
$\left\{B_{1},B_{2},\dots\right\}$ over the same set $N$, first, we
need to identify the correspondence between $A$s and $B$s and
re-order them. Then second, for each pair after re-ordered, the
dissimilarity of $A_{j}$ and $B_{j}$ is given by:
\begin{equation}
d_{j} =
\frac{\left|\left(A_j\cap\bar{B_j}\right)\cup\left(\bar{A_j}\cap
B_j\right)\right|}{\left|A_j\cup B_j\right|}.
\end{equation}
And the total dissimilarity can be calculated as
\begin{equation}
D = \frac{\sum_{i=1}^{K} d_{i}}{K}.
\end{equation}
Here, we use the dissimilarity function $D$ to quantify the
difference of different communities.

The paper is organized as following. The correlation between edge
weight and betweenness, and other properties has been studied in
Section \ref{correlat}. The negative results demonstrate that the
weight is an independent variable. Then in Section \ref{real}, we
compared community structures of weighted and corresponding
unweighted, disturbed weighted networks. The results demonstrate
that the weight has effects on communities, especially in dense
networks. In Section \ref{adhoc}, in order to show more results
about the effects of weight on communities, we investigated
idealized ad hoc weighted networks in detail. The results give us
systematic view about the effects of weight on community
structures. Finally we give some conclusion remarks.


\section{Community structures in Real and Unweighted, Inverse Weighted Networks}\label{real}

In this section, we focus on the effect of weight on community
structure in real weighted networks by comparing the communities
of real and its binary correspondence, and also inverse weighted
networks. Real networks include Econophysicists collaboration
network\cite{Fan}, BNU-email network and Rhesus monkey
network\cite{8}. For detecting and comparing community structure,
we take the largest connected cluster of the above networks. For
the Econophysicists collaboration network, it includes 271 nodes
and 371 edges. In order to distinguish the network with different
proportion of possible links, we define the denseness of network
as the ratio of existing links to the all possible links among the
nodes. The denseness for Econophysicists collaboration network is
0.01. The database for BNU-Email network includes the times of
Emails between any two mailboxes (*@bnu.edu.cn) in a week. The
network includes $740$ nodes and $1400$ links. We also use its
largest cluster, which includes $620$ nodes and $1117$ links. The
denseness of BNU-Email network is $0.006$. The monkey network
include 16 nodes and 69 edges. It is a connected network with
denseness equals $0.575$. So it is a relatively dense network.

As mentioned in the introduction, besides considering the
difference of suitable properties between unweighted and weighted
networks, an important way to investigate the effects of weight is
to study the impaction of weight redistribution to the network
properties. We have introduced the way to re-assign weights onto
edges with $p=1,-1$ for weighted networks \cite{Li}. Set $p=1$
represents the original weighted network given by the ordered
series of weights which gives the relation between weight and edge
but in a decreasing order,
\begin{equation}
W(p=1) = \left(w_{i_{1}j_{1}} = w^{1}\geq w_{i_{2}j_{2}} =
w^{2}\geq \cdots \geq
w_{\left(i_{L}\right)\left(j_{L}\right)}=w^{L}\right).
\end{equation}
$p=-1$ is defined as the inverse order as
\begin{equation}
W(p=-1) =\left(w_{i_{1}j_{1}} = w^{L}\leq \cdots \leq
w_{\left(i_{L-1}\right)\left(j_{L-1}\right)} = w^{2} \leq
w_{\left(i_{L}\right)\left(j_{L}\right)}=w^{1}\right),
\end{equation}
Here we only compare the community structures of original and
inverse weighted networks.

\begin{figure}
\center
\includegraphics[width=4.5cm]{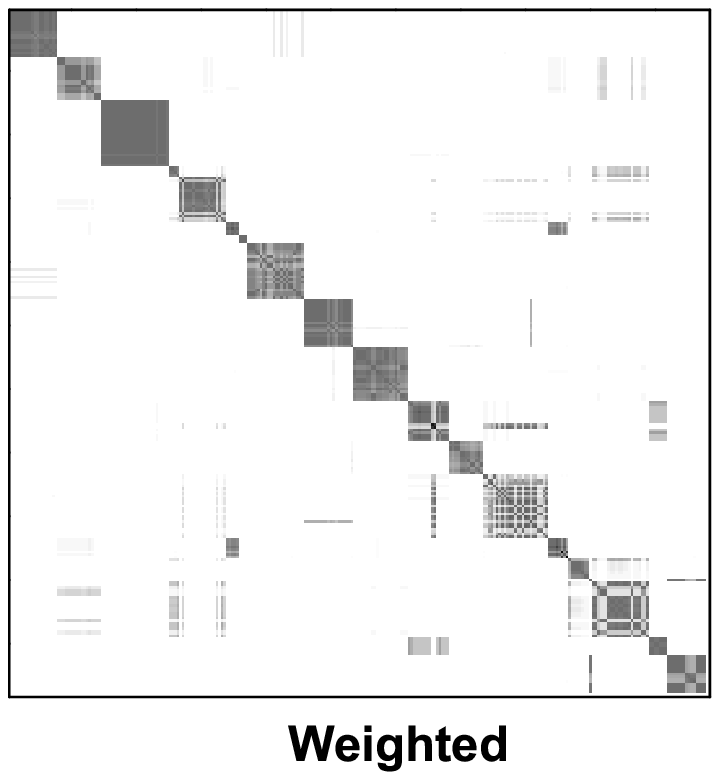}\includegraphics[width=4.5cm]{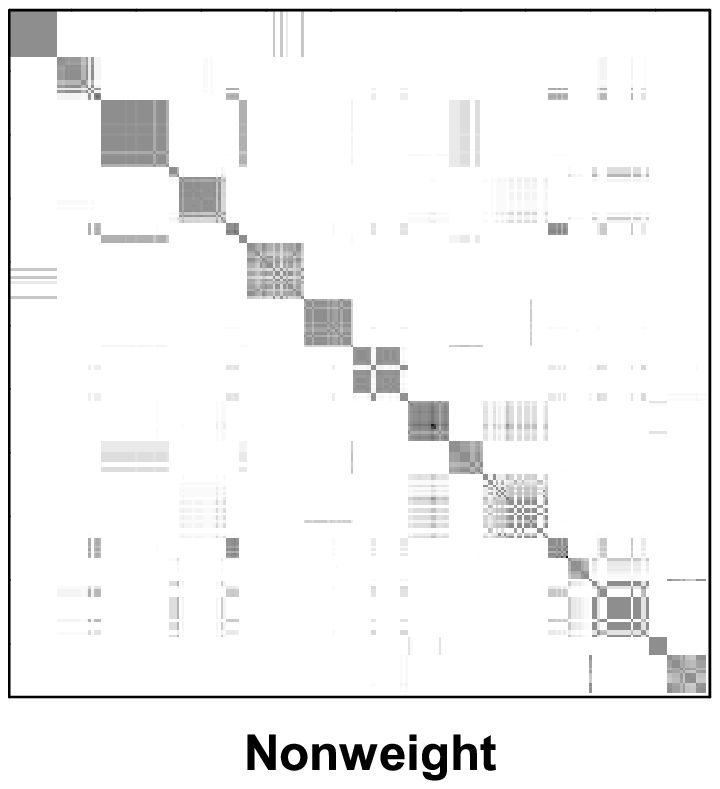}\includegraphics[width=4.5cm]{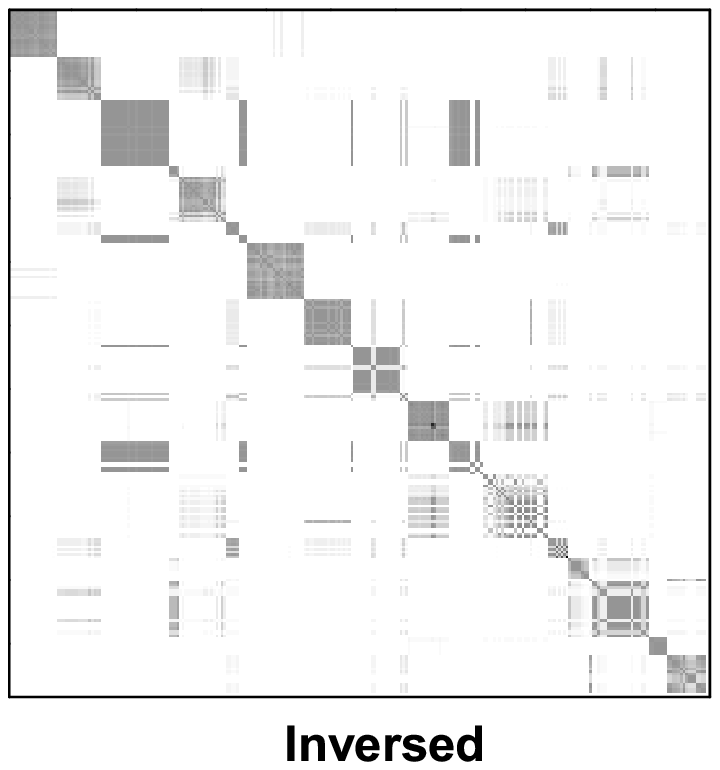}
\caption{The normalized and colored co-appearance matrix for (a)
original and (b) corresponding binary, and (c) inverse weighted
Econophysicists collaboration networks.} \label{e}
\end{figure}

We apply WEO algorithm 20 times for each network. Then the
community structure is shown by the corresponding co-appearance
matrix. Matrix ordering taken from a realization of WEO algorithm
in original network. In order to show the difference between
groups of original and unweighted, inverse weighted networks, we
keep the same order as the original one in the co-appearance
matrix of the unweighted and inverse weighted networks. Each
matrix gives the fraction of nodes classified in the same
partition over 20 realizations of WEO algorithm. The color of the
position $(i,j)$ corresponds to the fraction of times that nodes
$i$ and $j$ belong to the same group. Then the final communities
are given by the most probable partitions. The difference between
any pair of partitions is given by the dissimilarity function $D$.
In table 2, we show the comparison of the communities which formed
in original and unweighted and inverse weighted networks (with
$p=1,-1$ respectively).

\begin{center}
Table 2. Dissimilarities between communities of original and
binary, inverse weighted networks
\end{center}
\begin{center}
\begin{tabular}{c}
\begin{tabular}{|c|c|c|c|} \hline
& EP-SCN & BNU-Email & Monkey \\
\hline
D(Original-Binary) & 0.37 & 0.58 & 0.67\\
\hline
D(Original-Inverse) & 0.46 & 0.69 & 0.75\\
\hline
\end{tabular}\\
\end{tabular}
\end{center}

\begin{figure}
\center
\includegraphics[width=4.5cm]{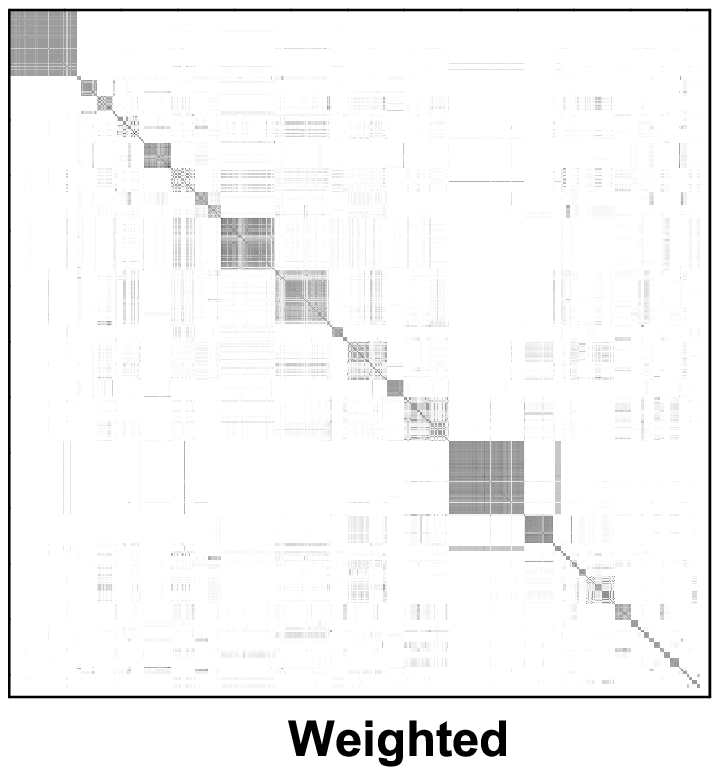}\includegraphics[width=4.5cm]{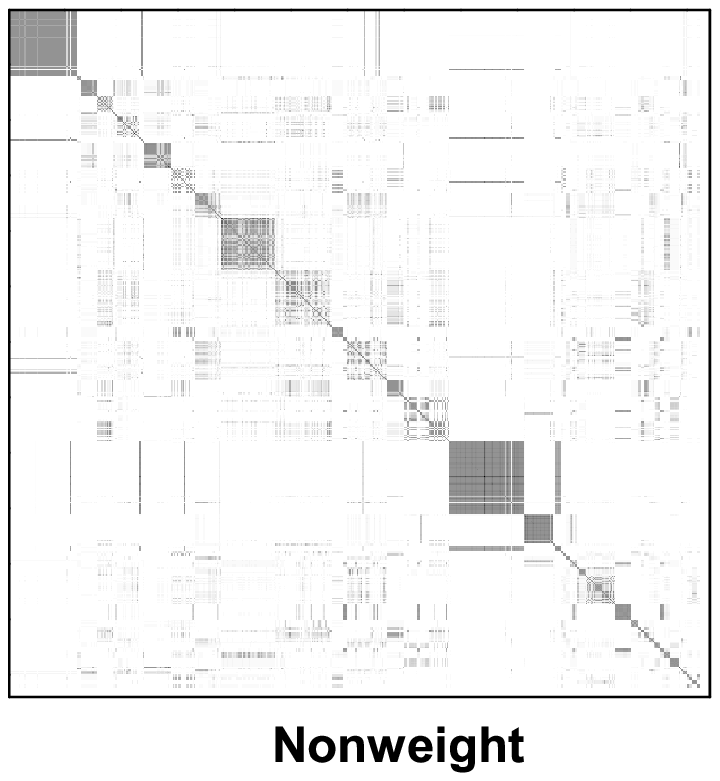}\includegraphics[width=4.5cm]{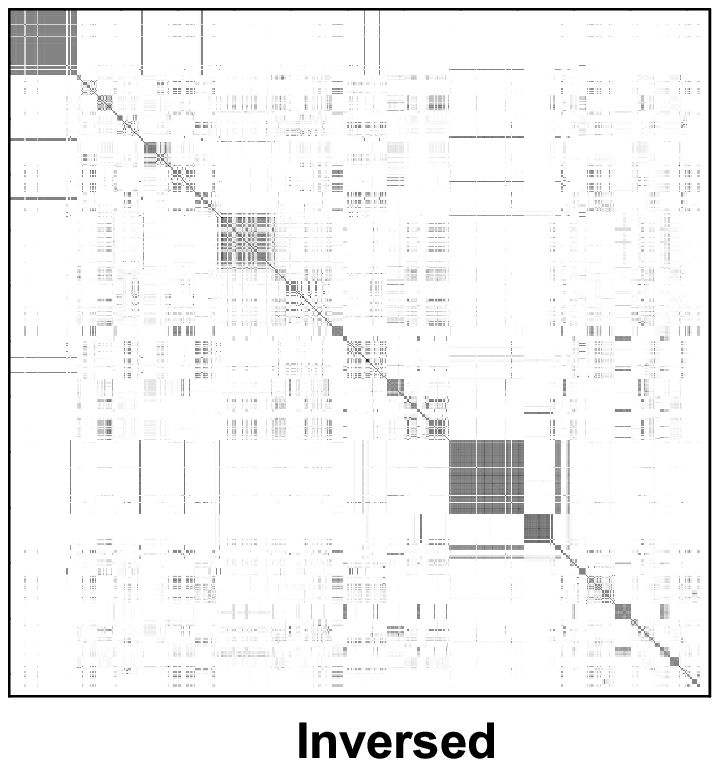}
\caption{The normalized and colored co-appearance matrix for (a)
original and (b) corresponding binary, (c) inverse weighted
BNU-email networks. }\label{d}
\end{figure}

\begin{figure}
\center
\includegraphics[width=4.5cm]{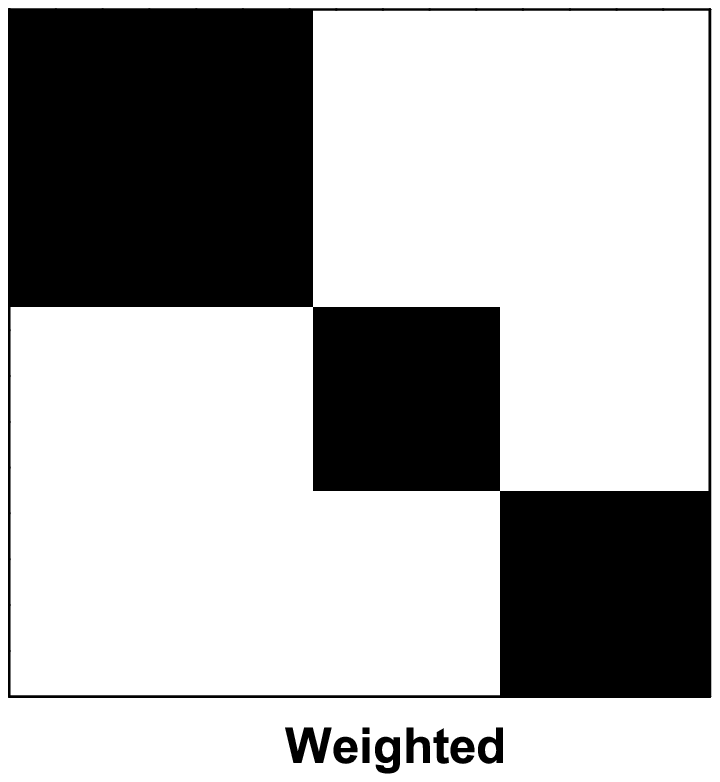}\includegraphics[width=4.5cm]{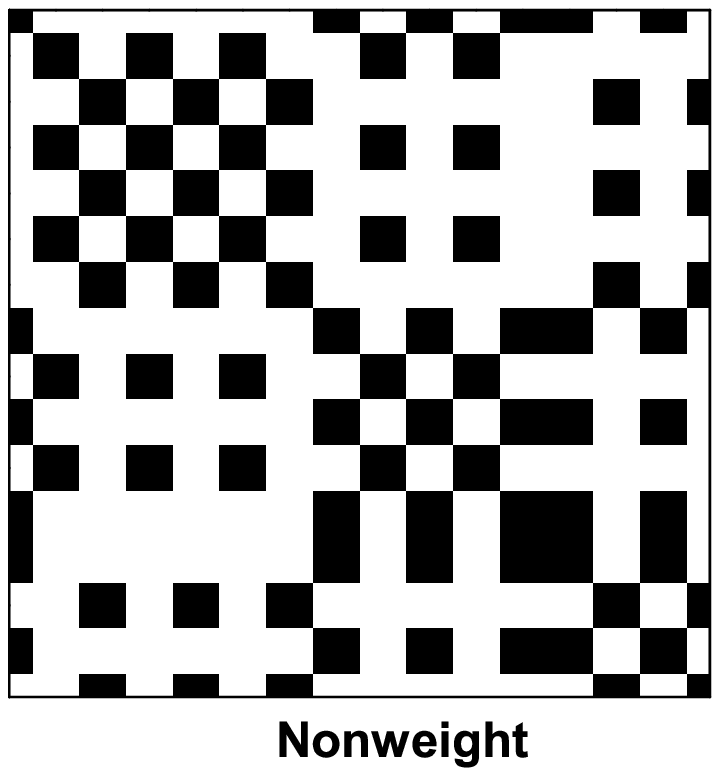}\includegraphics[width=4.5cm]{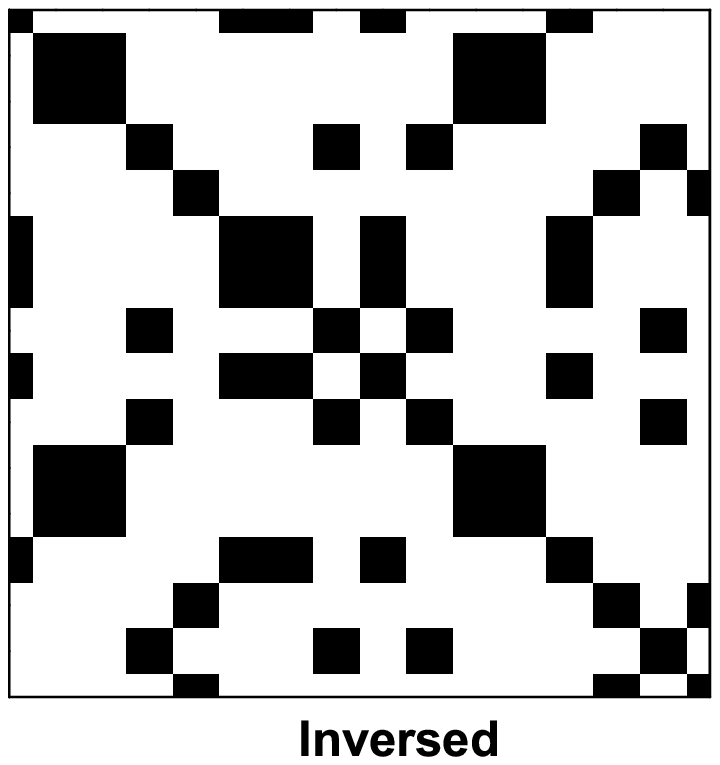}
\caption{The normalized and colored co-appearance matrix for (a)
original and (b) corresponding binary, (c) inverse weighted monkey
networks. The precision of the algorithm for original and binary,
inverse weighted networks are 1.00, 0.80, and 0.83 respectively.
The dissimilarity of the results are much larger than the other
two networks. It seems that weight plays more important role in
dense networks.}\label{f}
\end{figure}

As shown in the figures and Table 2, link weights indeed affect
the community structure which is related to the global structure
of networks. There are dissimilarities between partitions of
weighted and its binary correspondence networks, especially for
Rhesus monkey network. It seems that weight has more important
effects on dense networks.

\begin{figure}
\center
\includegraphics[width=7cm]{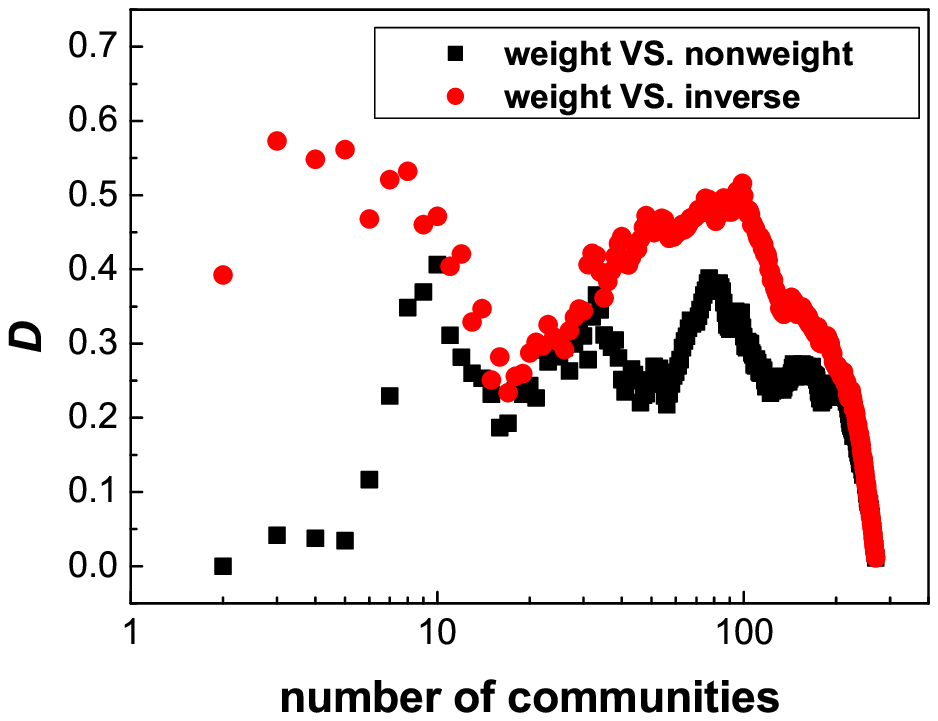}\includegraphics[width=7cm]{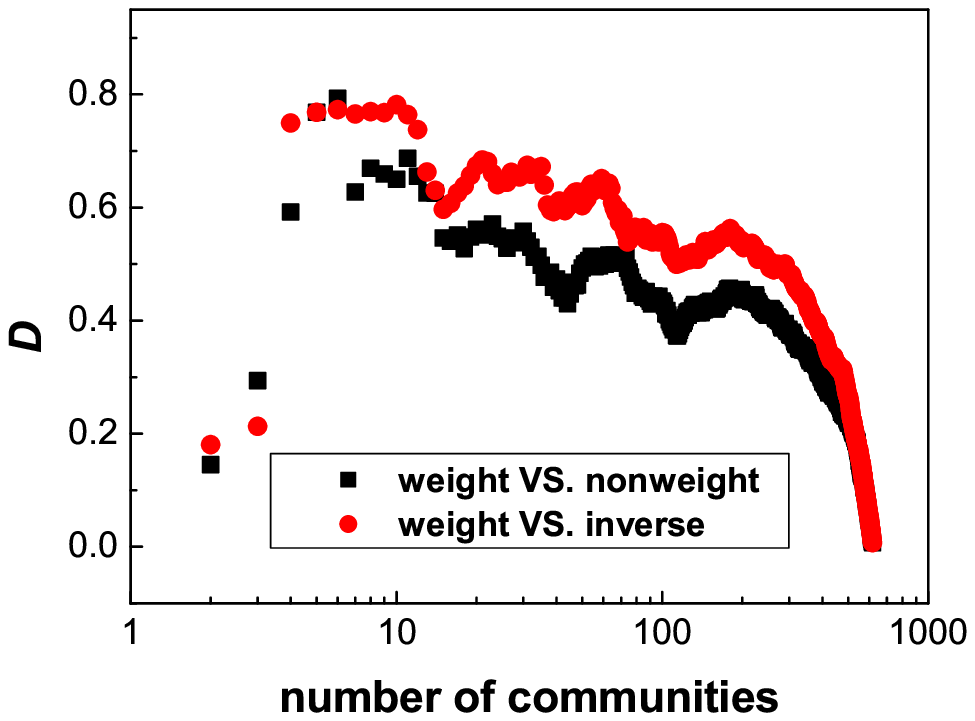}
\includegraphics[width=7cm]{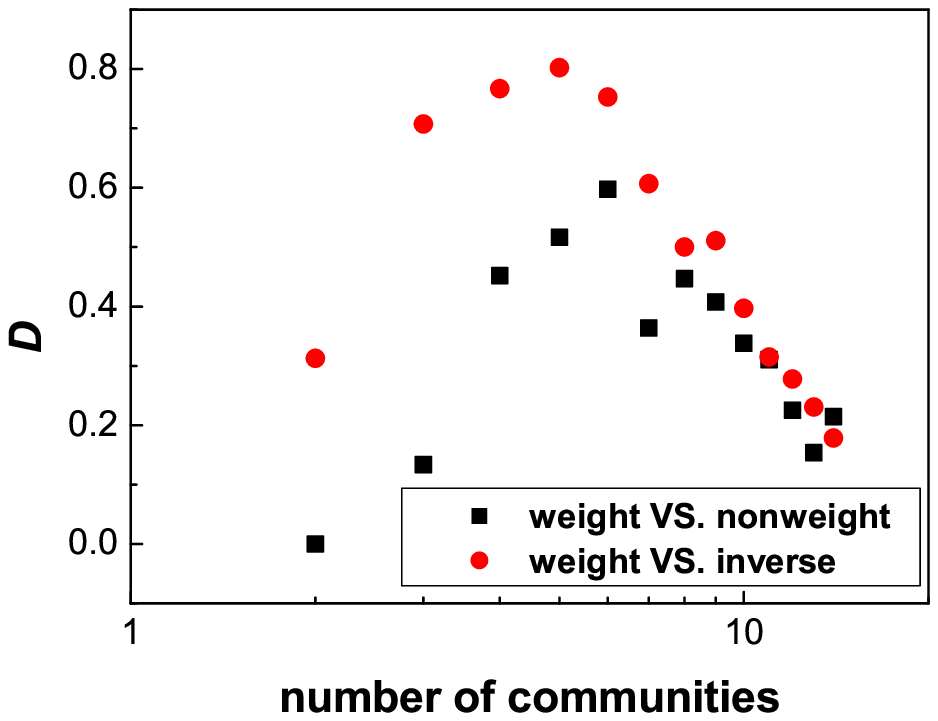}
\caption{A: The dissimilarity of communities gotten by GN
algorithm for original and binary, inverse weighted networks of
(a)Econophysicists collaboration network, (b)BNU-email network,
and (c)Monkey network. }\label{g}
\end{figure}

The WEO algorithm can only give us best partition of the network.
Using GN algorithm, we can compare the process of dividing networks
in to groups. Fig. \ref{g} shows the dissimilarity between the
results of original and binary, inverse weighted networks. It could
be found that the dissimilarity usually has the larger value when
the division of networks has gone into the inner structure of the
community. That also indicates the weight has bigger influence in
denser networks.

\

\section{Results on Idealized Weighted Networks}\label{adhoc}

Inspired by the above empirical studies, we constructed more
idealized \emph{ad hoc} weighted networks and try to get more
systematic conclusions about the effects of weight.

The idealized networks is firstly introduced by Newman and used by
many other authors\cite{PRE,compare}. Each network consists of
$n=128$ vertices divided into four groups of $32$ nodes. Vertices
are assigned to groups and are randomly connected to vertices of
the same group by an average of $\langle k_{intra}\rangle$ links
and to vertices of different groups by an average of $\langle
k_{inter}\rangle$ links. The average degree of all vertices are
fixed, namely $\langle k_{intra}\rangle+\langle
k_{inter}\rangle=16$. With $\langle k_{intra}\rangle$ increasing
from small, the communities become more and more diffuse, and it
becomes more and more difficult to detect the communities. For a
given network topology, here we assign similarity weight to each
link. The intragroup link weight is assigned as $w_{intra}$, while
the intergroup link weight is assigned as $w_{inter}$. In
practise, the relationship among the nodes in groups is usually
much closer than the relationship between groups. So $w_{inter}$
is normally less than $w_{intra}$. Similarly with $\langle
k_{intra}\rangle + \langle k_{inter}\rangle = 16$, we require
\begin{equation}
\langle w_{intra}\rangle + \langle w_{inter}\rangle = 2.
\end{equation}

We use \emph{ad hoc} networks with uniform distribution of link
weights here. For a given network topology with certain $\langle
k_{inter}\rangle$, weights are taken randomly from a $0.5$ interval
around $\langle w_{intra}\rangle$ and $\langle w_{inter}\rangle$
respectively for intragroup connections and intergroup connections,
that is $\langle w_{intra}\rangle-0.25, \langle
w_{intra}\rangle+0.25$ and $\langle w_{inter}\rangle-0.25, \langle
w_{inter}\rangle+0.25$ respectively. In the following simulations,
we take $\langle w_{intra}\rangle$=1.6 and so that $\langle
w_{inter}\rangle$=0.4.

\begin{figure}
\center
\includegraphics[width=7cm]{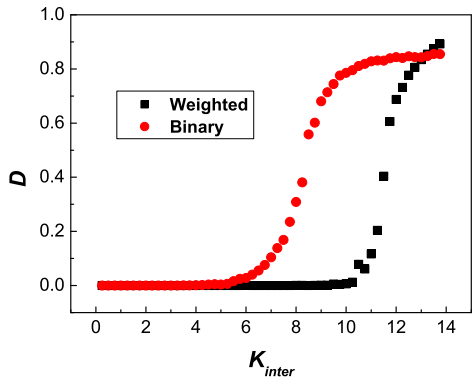}
\caption{The dissimilarity between groups found by WEO algorithm
and the presumed community structure. With the link weight
assigned by the method described in the text, weighted networks
have larger spectrum of $\langle k_{inter}\rangle$ than binary
networks.}\label{accuraBW}
\end{figure}

Now we exam the effects of weight on community structures based on
idealized weighted networks. First, we applied WEO method to the
binary and weighted \emph{ad hoc} networks. The groups gotten by the
algorithm are compared with the presumed communities with the
dissimilarity function $D$. We could found in Fig.\ref{accuraBW},
with the increasing of $\langle k_{inter}\rangle$, $D$ increase
sharply when $\langle k_{inter}\rangle$ is larger than 8 in binary
networks. But in weighted networks, the spectrum of $\langle
k_{inter}\rangle$ is much larger. Even when $\langle
k_{inter}\rangle$ is around 10, The network could grouped correctly.
For weighted networks, the community structure defined by weighted
modularity $Q^w$ integrate links with link weights. It shows that
the weight has an important effect on community structure.

\begin{figure}
\center
\includegraphics[width=7cm]{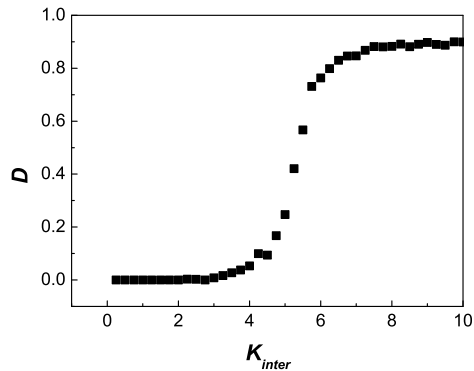}
\caption{The dissimilarity between groups in original networks and
the groups with random redistributed link weights. The communities
are found by WEO algorithm. With the increase of $\langle
k_{inter}\rangle$, link weights have more effects on community
structures.}\label{RandomD}
\end{figure}

Then we compared the community structure of weighted networks with
the groups after we inverse redistribute the link weight. The
community structures are all gotten by WEO approach. The original
idealized network is constructed with a given $\langle
k_{inter}\rangle$ and $\langle w_{intra}\rangle$=1.5. Then we
redistribute the link weights inversely according to the method
described in Section \ref{real}. Fig.\ref{RandomD} shows the
dissimilarity function $D$ between the original groups and groups
after disturbing link weights. The results are the average of 20
network realizations and 10 runs each. It is interesting to find
that $D$ is around 0 when $\langle k_{inter}\rangle$ is small but
it increases gradually with the increasing of $\langle
k_{inter}\rangle$. When $\langle k_{inter}\rangle$ is large
enough, it almost reaches 1. These results reveal that link and
link weight are two factors that determine the structure of
networks. When topological linkage dominates the structure of
networks, link weight plays less important role in networks. On
the other hand, in some networks, especially in dense networks
such as Rhesus monkey network, link weight is crucial to the
network structures.

\section{Concluding Remarks}\label{conclud}

In this paper, in order to investigate the role of weight, we pay
much attention to the influence of the weight to the results of
community structures. Besides the idealized ad hoc weighted
networks, the econophysicist collaboration network, Rhesus monkey
network, and BNU-email network are analyzed by using the WEO
algorithm. Using the dissimilarity function $D$ to measure the
difference of two kinds of community structures, we investigate
the different results of partition for non-weighted, weighted, and
inverse weighted networks. It is found that weight do have
influence on the formation of communities structure. That means:
1, the weight do have important role to the network structures; 2,
the community structure is a suitable global properties to reflect
the effect of weight. It has been also found that the weight is
more significant for dense networks. For a sparse network, the
existence or not of edges have bigger influence to community
structure of networks than the weight.

\end{document}